# Spatially resolved mass flux measurements with dual comb spectroscopy

David Yun[1,*], Ryan K. Cole[1], Nathan A. Malarich[1], Sean C. Coburn[1], Nazanin Hoghooghi[1], Jiwen Liu[2], Jacob J. France[3], Mark A. Hagenmaier[4], Kristin M. Rice[4], Jeffrey M. Donbar[4], Gregory B. Rieker[1,5]

[1]*Precision Laser Diagnostics Laboratory, Department of Mechanical Engineering, University of Colorado Boulder, Boulder, CO 80309, USA*
[2]*Taitech Scientific Solutions Incorporated, Beavercreek, OH 45430*
[3]*Innovative Scientific Solutions Incorporated, Dayton, OH 45459*
[4]*U.S. Air Force Research Laboratory, Wright-Patterson AFB, OH, 45433, USA*
[5]*email: greg.rieker@colorado.edu*
*\*Corresponding author: david.yun@colorado.edu*

**Abstract:** Providing an accurate, representative sample of mass flux across large open areas for atmospheric studies or the extreme conditions of a hypersonic engine is challenging for traditional intrusive or point-based sensors. Here, we demonstrate that laser absorption spectroscopy with mode-locked frequency combs can simultaneously measure all of the components of mass flux (velocity, temperature, pressure, and species mole fraction) with low uncertainty, spatial resolution corresponding to the laser line of sight, and no supplemental sensor readings. The low uncertainty is provided by the broad spectral bandwidth, high resolution, and extremely well-known and controlled frequency axis of stabilized, mode-locked frequency combs. We demonstrate these capabilities using dual frequency comb spectroscopy (DCS) in the isolator of a ground-test supersonic propulsion engine at Wright-Patterson Air Force Base. The mass flux measurements are consistent within 3.6% of the facility-level engine air supply values. A vertical scan of the laser beams in the isolator measures the spatially resolved mass flux, which is compared with computational fluid dynamics simulations. A rigorous uncertainty analysis demonstrates a instrument uncertainty of ~0.4%, and total uncertainty (including non-instrument sources) of ~7% for mass flux measurements. These measurements demonstrate DCS with mode-locked frequency combs as a low-uncertainty mass flux sensor for a variety of applications.

## 1.     Introduction

Accurate measurements of gaseous mass flux, the total flow rate of molecules through a volume, are essential to many research fields. For example, species mass flux measurements have been used to study greenhouse gas emissions of cities and arctic environments [1–6], evapotranspiration of forests and agricultural sites [7–9], and atmospheric effects of wildfires [10,11]. Mass flux measurements are also useful for a host of engineering applications, such as aeropropulsion research, where air mass flux is essential for analyzing inlet phenomena in hypersonic engines [12], determining flow enthalpies in arc-jet facilities [13], and characterizing impulse in rotating detonation engines [14]. However, direct measurements are extremely challenging in large-scale open environments and environments with extreme flow phenomena because it is difficult to accurately measure all of the components of mass flux (velocity and density, the latter being derived from temperature, pressure, and mixture composition). Here, we demonstrate non-intrusive, absolute measurements of mass flux in such environments with low uncertainty by leveraging the unique properties of stabilized, mode-locked frequency combs. In particular, the extremely precise and well-known spacing between the optical 'teeth' of the comb creates a near-perfect ruler against which to measure the velocity-dependent Doppler shift and the pressure-dependent widths of molecular absorption features.

The broad optical bandwidth of the combs enables measurement of a multitude of absorption features whose relative intensity collectively define the temperature of the gas. The optical bandwidth also enables accurate measurement of the gas composition, including the potential to resolve multiple gas species at once.

We use frequency combs to measure the mass flux in a direct-connect, ground-test, dual-mode ramjet propulsion system at the Air Force Research Laboratory. Scramjets/ramjets are air-breathing engines which typically operate at speeds above Mach 5. While short duration scramjet flights have been demonstrated via programs such as X-51 [15] and HIFiRE Flight 2 [16], achieving consistent operation requires improved understanding and optimization of engine designs. Specifically, complex flow patterns due to shock-boundary interactions around the supersonic inlet of flight vehicles (see Figure 1a) can lead to uncertainty in the inlet air mass flux. This in turn requires overdesign of flight surfaces and inlet properties to accommodate these uncertainties, at the expense of flight performance.

We determine the air mass flux with low uncertainty by measuring velocity, temperature, pressure, and species mole fraction through absorption spectroscopy of hundreds of water vapor quantum transitions in the $6880 – 7186$ cm$^{-1}$ spectral region ($1392 – 1453$ nm) with ~46,000 optical frequencies (comb teeth) of a stabilized all-fiber mode-locked dual frequency comb spectrometer. We take several measurements at varying flow conditions at the centerline of the dual-mode ramjet. This choice of facility allows for comparisons to both computational fluid dynamics (CFD) calculations and upstream facility flow rate measurements which are expected to have high accuracy as the facility is a closed, direct-connect system. Additionally, we perform a vertical scan at a single flow condition to within 1 mm from the wall to provide a mass flux spatial profile. The measured flow properties and resulting air mass flux results are compared to facility measurements and computational fluid dynamics (CFD) calculations across a range of flow conditions. Through a detailed analysis, we estimate 0.4% instrument uncertainty in the measured mass flux. Additional systematic uncertainties bring the overall uncertainty to ~7%, largely due to uncertainty in the absorption model that is used to interpret the instrument measurements (and which can be improved in the future).

This is the first demonstration of dual comb absorption spectroscopy (DCS) measurements of air mass flux, and the first demonstration of DCS in a hypersonic aeropropulsion environment. Our results demonstrate the potential of broadband dual-comb spectroscopy for accurate, precise, and nonintrusive mass flux measurements in high-speed flows. While the DCS system used here is tailored for water-air mass flux and is thus based on $H_2O$ absorption in the near-infrared spectral region, the mode-locked frequency comb spectral range can be tailored to be applicable to other gas mixtures and encompass measurements of more species relevant to greenhouse gas monitoring and combustion [17–23]. Additionally, works such as [24–28] have demonstrated that DCS instrument measurements can be easily expanded to large areas and open environments, enabling mass flux measurements for large-scale studies in environmental, atmospheric, and agricultural research. Recent work on chip-sized frequency combs [29–31] and electro-optic frequency combs will further open mass flux measurement possibilities for DCS [32–34].

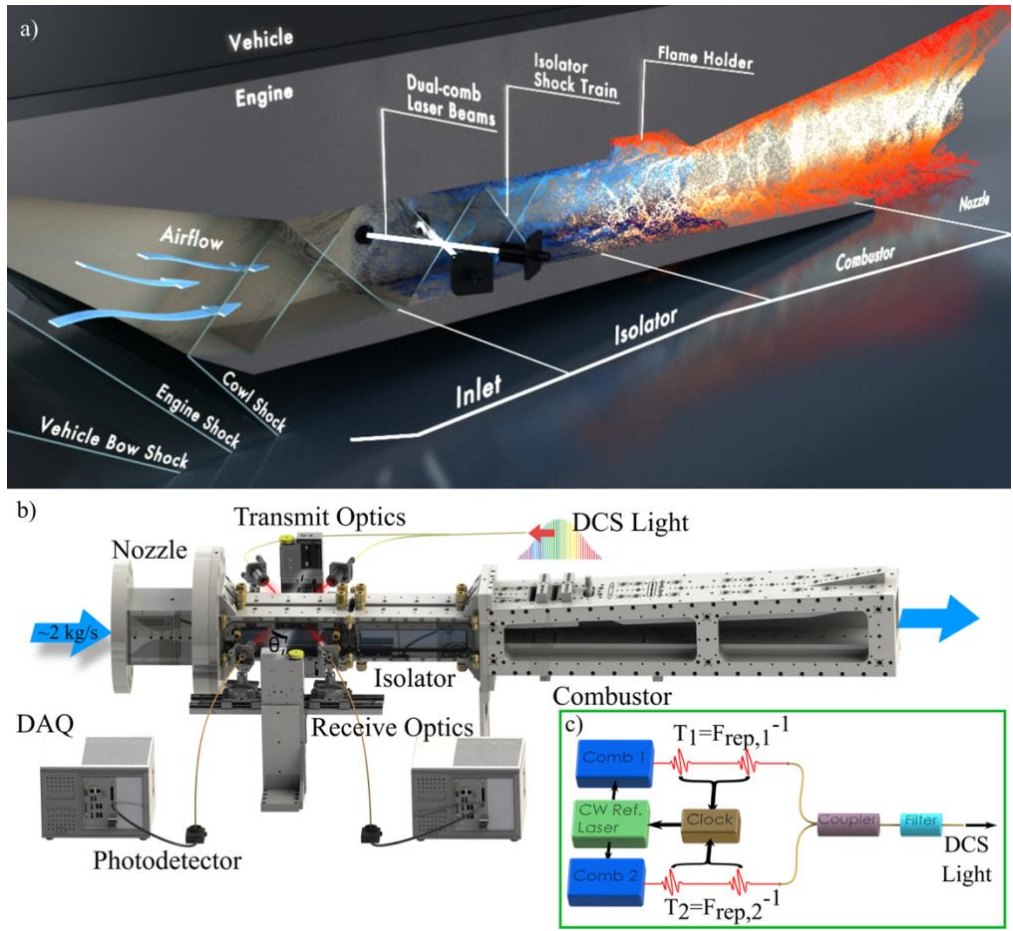

Fig. 1. DCS measurements in a dual-mode ramjet test engine. (a) The main components of a ramjet/scramjet are the inlet and isolator which compress the air via an oblique shock train, the combustor which injects and ignites fuel to heat the air, and a nozzle which expands and accelerates the air creating thrust. (b) Light from the frequency combs launch into the isolator at angle θ. Resulting light is caught onto multi-mode fiber and detected on a pair of fast photodetectors via a high-speed field-programmable gate array data acquisition system. Pitch and catch optics are placed on translation stages to allow for measurements at different heights. (c) DCS consist of two frequency combs whose repetition rates are read by a highly accurate clock and tightly controlled by a reference laser. Light from both combs is combined via a coupler and filtered to the $H_2O$ infrared absorption band.

## 2. Dual frequency comb spectroscopy for air mass flux measurements

### 2.1 Optical mass flux measurement

Mass flux is the product of the velocity ($U$) and density ($\rho$) in a flow.

$$\dot{m} = \rho U \qquad (1)$$

If we assume that the composition of the gas besides $H_2O$ (the molecular absorption target for the present measurement) is dry air, we can determine the density via the ideal gas law using temperature, pressure, and $H_2O$ mole fraction and ultimately calculate mass flux with the following equation:

$$\dot{m} = \frac{(\chi_{H2O}M_{H2O} + (1-\chi_{H2O})M_{air})PU}{RT} \qquad (2)$$

Here $M_{H2O}$ and $M_{air}$ are the molar masses of water and dry-air, and R is the gas constant. The rest of the terms: pressure ($P$), temperature ($T$), H$_2$O mole fraction ($\chi_{H_2O}$), and velocity ($U$), can be extracted from laser absorption spectra measured with an appropriate optical configuration and laser.

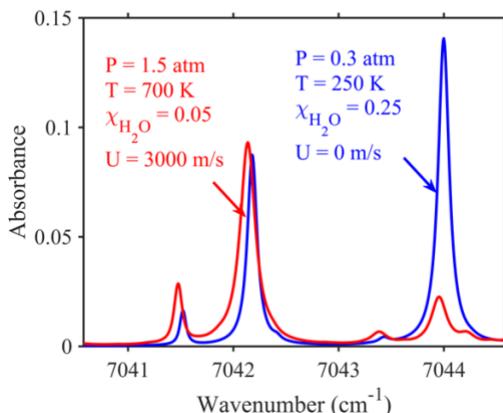

Fig. 2. H$_2$O absorption features at two different thermodynamic conditions: P = 0.3 atm, T = 250 K, $\chi_{H_2O}$ = 0.25, U = 0 m/s (blue) and P = 1.5 atm, T = 700 K, $\chi_{H_2O}$ = 0.05, U = 3000 m/s. Here, the Doppler shift is associated with a beam angle of 35° to the normal of the flow direction.

DCS (and laser absorption spectroscopy in general) measures laser absorption at frequencies resonant with the rovibrational quantum state transitions of a constituent molecule in a gas sample or flow. Figure 2 shows a small portion of the H$_2$O absorption spectrum to demonstrate the large differences in the spectrum that are induced by different velocity, pressure, temperature, and species mole fraction. Flow velocity induces a Doppler shift in the absorption feature positions if the laser beam is angled upstream or downstream to the bulk gas flow [35]. In the present measurement, we configure the two beams in a crossed configuration with one beam angled upstream and the other downstream with equal angles (see Fig .1a). In this case, equal and opposite Doppler shifts are induced in the absorption spectra measured with each beam. The width of the absorption features is proportional to pressure. The magnitude of the absorption features is proportional to the species mole fraction. Finally, the relative intensities between the features in the spectrum are dependent on temperature because absorption transitions arise from quantum states with different lower state energies. Thus, pressure, species mole fraction, velocity, and temperature of the probed region of the flow can be determined by comparing the size, shape, and position of measured absorption features with an absorption model of the expected spectra for a given set of conditions.

*2.2    Dual frequency comb spectroscopy*

Mode-locked frequency comb lasers have emerged as powerful sources for absorption spectroscopy [36,37], which can be used to simultaneously measure all of the components of mass flux. Frequency combs emit a train of optical pulses comprising multitudes of evenly spaced, discrete optical frequencies (referred to as "comb teeth") across a broad spectral bandwidth. The optical frequencies of the comb teeth are related through Eq. 3:

$$f_n = nf_{rep} + f_0 \qquad (3)$$

where $f_0$ is the carrier envelope offset frequency, $n$ is the comb tooth number, and $f_{rep}$ is the pulse repetition rate of the laser. Thus, the absolute frequency of each individual comb tooth is set with only two degrees of freedom ($f_{rep}$ and $f_0$). These parameters can be precisely measured and controlled (as described below) to create a near-perfect optical ruler with which to measure the position, shape, and magnitude of absorption features.

A key challenge in spectroscopy with frequency combs is preserving the resolution and tight control of the fine comb tooth lines in the detection of the absorption information contained in the thousands of individual optical frequencies after they have traveled through the environment of interest. We address this challenge using dual comb spectroscopy (DCS), in which two frequency combs are coherently locked to one another with slightly different comb tooth spacing. The comb tooth spacing is set so that successive pairs of teeth from the two combs have a unique frequency offset from each other that gives rise to a unique heterodyne beat signal at the difference frequency between the teeth. Thus when the combs are interfered on a photodetector, we down-convert pairs of comb teeth from their optical frequencies to corresponding heterodyne beat signals at easily detectable radio frequencies [38]. The attenuation of the comb teeth by absorption is reflected in the strength of the heterodyne beat signals, and thus the absorption spectrum can be read out on a tooth-by-tooth basis maintaining the frequency accuracy and precision of the combs themselves. The detection requires a single fast photodetector to capture the interference signal (the interferogram), which yields the laser transmission spectrum after applying a Fourier transform. Each interferogram is produced at a rate equal to the difference in $\boldsymbol{f_{rep}}$ between the two combs, allowing for rapid acquisition of individual transmission spectra (1.4 kHz here). As noted later, many of these spectra can be averaged together to achieve high levels of precision due to the stable, coherent nature of the frequency combs.

The frequency combs used here are Erbium-doped, polarization-maintaining-fiber based frequency combs with a pulse repetition rate of 200 MHz [39]. Figure 1c shows an overview of the DCS configuration. The $f_0$ of each frequency comb is controlled and stabilized using the $f$-to-$2f$ locking scheme [40–42]. In this approach, each comb is broadened across an octave of optical frequency space, and a high frequency tooth is frequency doubled and interfered with a lower frequency tooth from the same comb. In this scheme, $f_0$ is determined by the heterodyne beat frequency between the two interfered teeth which we stabilize at a specified RF frequency through feedback to the frequency comb oscillator pump current. Additionally, a tooth from each frequency comb is phase-locked to a common 2 kHz linewidth continuous wave (CW) reference laser. This establishes mutual coherence between the two comb sources, and provides the means to stabilize the second degree of freedom ($f_{rep}$) of each comb. Specifically, the heterodyne beat frequency between the CW laser and the tooth from each laser is measured with a frequency counter and stabilized through feedback to high-speed piezoelectric transducers glued to the fibers of each laser cavity. The repetition rate for each laser is also continuously measured using a frequency counter, and we use slow feedback to the CW laser current (which controls the CW laser wavelength) to maintain stable $f_{rep}$ [43].

Using this stabilization approach, the quality of the optical ruler for measuring the absorption spectrum is set by the stability and accuracy of the frequency reference for the frequency counters, and the amount of drift in $f_{rep}$ allowed by the CW laser feedback. In this study the frequency reference is a simple ovenized quartz oscillator (Wenzel 501-27514-21) with a relative frequency accuracy of 1.5 ppm. The CW laser digital feedback loop allows a drift in $f_{rep}$ of less than 25 ppb. Thus, the frequency reference is the limiting factor in the ultimate accuracy of the comb tooth frequencies. Still, the 1.5 ppm relative accuracy of the frequency reference enables the spacing of the comb teeth to be known to within ~300 Hz ($1.0 \times 10^{-8}$ cm$^{-1}$) at all times and the absolute position of teeth to within 300 MHz (0.011 cm$^{-1}$). This leads to low instrument uncertainty, as described later. If further accuracy is required the frequency reference can be replaced with a GPS-disciplined oscillator which has a relative accuracy of 10 ppt, and the feedback to the CW laser can be improved by switching to a high-speed analog control feedback on the CW laser to stabilize $f_{rep}$.

With the dual-comb spectrometer employed in this study, we achieve ~300 cm$^{-1}$ optical bandwidth encompassing hundreds of $H_2O$ absorption features with a spectral point spacing of 0.0067 cm$^{-1}$. These mode-locked dual frequency comb spectroscopy performance metrics thus blend broad bandwidth with very high resolution and frequency referencing that far exceeds most traditional absorption techniques. For example, a TDLAS measurement with a 15 cm etalon reference would need real-time knowledge and control of the etalon length to 225 nm to achieve the frequency referencing performance of the dual comb. DCS also allows for fast acquisition. In this work, each spectrum is acquired in 0.7 ms, and the stability of the lasers allow for long averaging times to optimize SNR and precision as needed (in this case 60 s of data is averaged to reach <0.33% precision across all measurements).

*2.3    Instrument uncertainty*

The DCS performance characteristics relate directly to low *instrument* uncertainty for each of the measured components of mass flux. We define *instrument* uncertainty to be the fundamental limit of the measurement uncertainty imparted by the stability and accuracy of the DCS. For velocity measurements, the absolute frequency error of the frequency combs cancel in the crossed beam configuration (see Supplement 1 Sec. 1), and only uncertainty in the comb teeth spacing imparts an instrument uncertainty. As stated above, this is controlled to 0.00015%, and sets the lower limit of achievable uncertainty in the measurement of velocity. The uncertainty of the comb tooth spacing also defines the lower limit of the achievable instrument uncertainty for pressure measurements (which are based on the relative width of the measured absorption features). We found that for pressure, temperature, and species mole fraction retrievals, uncertainty in the frequency axis also leads to additional error during the fit of the absorption model to the measurement due to correlation between the parameters (since all are retrieved simultaneously). As described in Supplement 1 Sec. 3.1 these uncertainties are still very low; 0.012% for pressure, 0.015% for species mole fraction, and 0.0042% for temperature.

While difficult to quantify, the DCS bandwidth, resolution, and lack of instrument lineshape (distortion) also help reduce uncertainty in the measurement [44]. By incorporating many features, scatter in the absorption model error from feature to feature are averaged out. Additionally, the broad bandwidth increases robustness against optical interference effects such as etalons and ambiguities in laser intensity baseline which can distort feature intensity retrievals [45]. For temperature, incorporating a multitude of features from different lower state energies offers improved dynamic range and reduced uncertainty with respect to traditional tunable laser two-line thermometry [46].

Altogether, mode-locked DCS is therefore well-suited for mass flux measurements due to the low levels of uncertainty that are achievable through the quality of the optical spectrum reference, its broad bandwidth and high resolution, and point spacing accuracy. We describe later that background absorption, noise, absorption model error, laser angle uncertainty, and other systematic uncertainties outside of the DCS instrument itself increase the overall uncertainty. However, many of these systematic uncertainties can continue to be improved through changes to the optical setup and better absorption models.

**3.    DCS experiment in a direct-connect dual-mode ramjet engine**

We demonstrate dual comb mass flux sensing in a continuous-flow, direct-connect, supersonic combustion research facility at Wright-Patterson Air Force Base [47]. The test article is a ground-test dual-mode ramjet. A dual-mode ramjet can operate in either ramjet or scramjet mode. Ramjets and scramjets are air-breathing engines which typically operate at speeds above Mach 5 and slow and compress air with a series of shock waves as opposed to the fans and turbomachinery common to lower-speed engines. Scramjets differ from ramjets in that air in the engine never slows below the speed of sound allowing for faster flight speeds. Figure 1a shows a schematic of a dual-mode ramjet. During supersonic flight, air passes through shockwaves emanating from the leading edges of the aircraft. It then enters the inlet/isolator region and is further compressed by a series of shock waves that raises the pressure and temperature of the air, while reducing its velocity. Fuel is then mixed with the air and ignited in the combustor. The flow finally expands through a nozzle to provide thrust. Air mass flux through the engine is a critical parameter. Taken together with fuel flow rate and thrust, the overall performance of the engine can be assessed. It is a difficult parameter to measure, as intrusive diagnostics cause performance-changing perturbations in supersonic environments. It is also a difficult parameter to calculate for 'free-jet' ground-test facilities (where the test article is placed inside a supersonic wind tunnel), or flight engines, where there is high uncertainty regarding the air flow into versus around the engine.

The isolator of a direct-connect test facility is ideal for this initial demonstration because the mass flux can be determined with reasonable accuracy based on the test facility air flow measurements (all of the air flow from the facility compressors are directed through the test article in a direct-connect facility). The test article is 10.2 cm wide, 3.8 cm tall, and 61 cm long with optical access via 2.1-cm-thick quartz windows. See Supplement 1 Sec. 4 for a more detailed description of the dual-mode ramjet test model.

We place the dual-comb spectrometer in a control room adjacent to the test cell. A coupler combines light from each of the two combs and sends the combined light onto two separate single-mode fiber paths. The light from each path then passes through free-space optical filters and back into single-mode fiber for transmission to the test cell. In the test cell, the light is collimated and passed through the isolator windows. We set both optical paths parallel to the bottom of the isolator. Using high precision rotation mounts we angle one beam in an upstream-propagating direction and the other in a downstream-propagating direction (hereafter referred to as the upstream and downstream beam respectively) at angles of 35° and -35° ($\pm$0.25°, see Supplement 1, Sec. 3.3) respectively from the normal of the isolator windows, as shown in Figure 1b. After passing through the isolator, the light is focused into multimode optical fibers via 1 cm diameter optics that are robust to the intense turbulence-induced beam steering that occurs through the test section. The transmit and receive optics rest on vertical translation stages that enable measurements at different heights in the isolator. The DCS signals are measured by photodetectors and recorded on two separate 250 MS/s digitizers, which are both clocked by the pulse repetition rate of one of the combs.

We obtain interferograms resulting in spectra between 6880 – 7186 cm$^{-1}$ (1392 – 1453 nm) using 46,000 comb teeth (each with ~25 kHz linewidth) at a rate of 1.4 kHz (Figure 3). This region spans hundreds of observable $H_2O$ absorption features. Phase slip and timing jitter between interferograms due to small fluctuations in the fibers and optics after the light leaves the combs but before the two combs are combined are corrected in post-processing to allow for coherent averaging [48], which we do for 60 seconds for each measurement to maximize SNR and minimize statistical fit uncertainties. Figure 3a shows the transmission spectra from the highest velocity condition (run 5). We fit the averaged measured spectra using the Levenberg-Marquardt non-linear least-squares method with a modified free induction decay approach. The specifics of this technique is given in more detail in Ref. [45] as well as in Supplement 1 Sec. 2. Briefly, cepstral analysis largely separates the molecular response from baseline effects in the time domain. The portion of the signal that is dominated by the molecular response is fit with an absorption model to extract the best-fit parameters. The fits use speed-dependent Voigt absorption models created from the high-temperature $H_2O$ absorption database developed by Schroeder et al. [49], which improves upon HITRAN2012 [50] especially for the high-temperature conditions present in the isolator. An example model generated at the best-fit conditions is compared with the data in Figure 3d. All open-air optical paths outside of the test engine are purged with dry air to reduce interfering background absorption from ambient water vapor. Remaining background absorption is measured before and after each ramjet run with cold dry air flowing through the isolator and subtracted from the operational run data as described in Supplement 1 Sec. 3.4.

The mass flux measurement is localized to the vertical and axial (streamwise) location of the crossed lasers. For the rectangular isolator duct, we can assume a symmetrical flow along the transverse axis meaning the crossed beams are able to represent a transverse-averaged slab with a 7 cm streamwise due to the angling of the lasers and a 1 mm height resolution corresponding to the beam diameter. In this study, we demonstrate spatially resolved measurements in the vertical direction by scanning the crossed beams vertically. The flow conditions tested here were designed to be simple (i.e. no distortion generator or isolator shock train), so that the DCS measurements could be readily compared with area-normalized facility mass flow rate measurements and CFD calculations. This resulted in minimal gradients along the axial direction of the isolator. Future measurements of more complex conditions could include translation of the laser beams in the axial direction to achieve a 2D measurement of the flowfield.

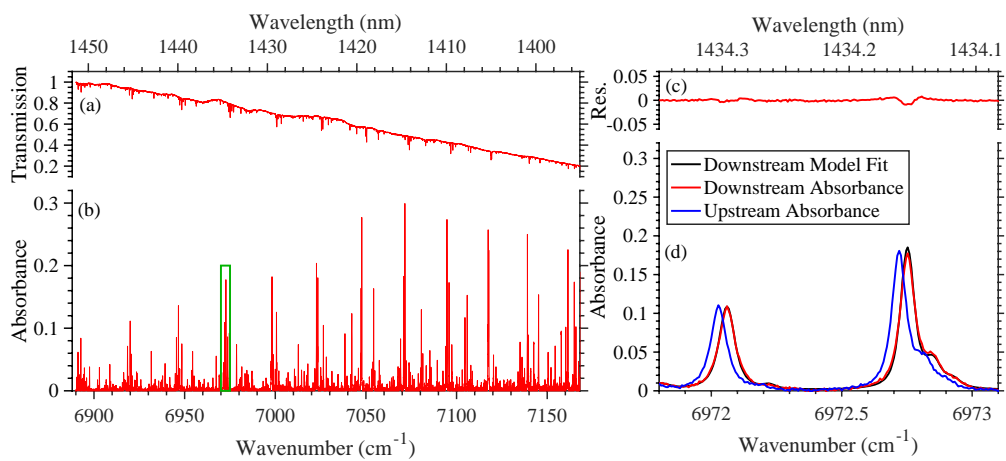

Fig. 3. Measured spectra from run 5 with 60-second averaging. Panel a) shows the transmission spectrum from the downstream-propagating beam. Panel b) shows the corresponding absorbance spectra with laser intensity baseline and background absorbance removed. Panel c) shows the fitted residual (data-model) of.Panel d) which shows a zoom view of the absorbance spectrum corresponding to the green box in panel b) together with the fitted model and the upstream-propagating beam spectrum.

## 4. Results and CFD comparison

### 4.1. Centerline measurements

In the first test configuration, we perform measurements at the centerline of the isolator (19 mm from the bottom) while the facility parameters are varied to produce different velocity, temperature, and pressure conditions (tabulated in Supplement 1 Sec. 5). The DCS measurement results for velocity, temperature, pressure, $\chi_{H2O}$, and the consequent total mass flux calculations are shown in Figure 4.

The simple flow conditions in this study feature a uniform core flow with relatively small boundary layers. Thus, a centerline measurement can be reasonably compared against area-normalized facility mass flow rate measurements. We chose a direct-connect facility (where the entire mass flow of the facility compressors must pass through the engine) so that the compressor mass flux could serve as a direct comparison with the measurements. The facility mass flow rate measurements are taken with flow meters upstream of the isolator which are expected to have low uncertainties. We normalize the flow rate with the isolator cross-sectional area to provide a facility-derived mass flux, which is shown in Figure 4. The measurements are within 0.1% to 3.6% of the facility-derived mass flux across all tests. As discussed later, we estimate the DCS mass flux uncertainty to be 7%. The mass flux measurements thus agree within their uncertainty to the facility values, suggesting that the uncertainty estimate is correct (if not conservative).

The facility measurements only enable mass flux comparisons, so we use a 3D CFD flow solution for comparison with the other measured parameters (velocity, temperature, species mole fraction and pressure). The simulations are performed in CFD++ (Metacomp Technologies, Inc.) based on a full 3D Reynolds-Averaged Navier Stokes approach with the two-equation cubic $k - \varepsilon$ turbulence model and an isothermal wall condition. To compare the CFD and DCS values, we simulate a line-of-sight absorption measurement through the CFD at the same location as the real-world measurement [51]. We extract flow properties voxel-by-voxel along a simulated beam path through the CFD to produce an integrated absorption that accounts for nonuniformities in flow parameters. CFD-derived flow values are then extracted from the simulated integrated spectra using the same fitting algorithm used to fit the DCS spectra. This process minimizes any differences between DCS and CFD values due to nonuniformity in the flow. The CFD-derived values for all of the measured parameters are shown in Figure 4. The CFD-derived parameters track the DCS-measured parameters. The CFD-derived velocities are within 2.6% of the DCS measured values. The CFD-derived temperature differs most from the DCS temperature at run 5 (4.8%) while differing less than 4% for other runs. The CFD-derived $\chi_{H2O}$ and pressure values exhibit the largest deviations from the DCS measured values, up to 6.1% and 7.3%, respectively. These differences result in a total difference between the CFD-derived and DCS-measured mass flux ranging between 3.1% and 8.8%. We note that the CFD-derived values fall farther from the facility-measured mass flux than the DCS measurements for all conditions. We show in the next section that spatially resolved measurements are helpful in assessing how different CFD boundary conditions influence the ultimate CFD result, and can help improve the CFD results.

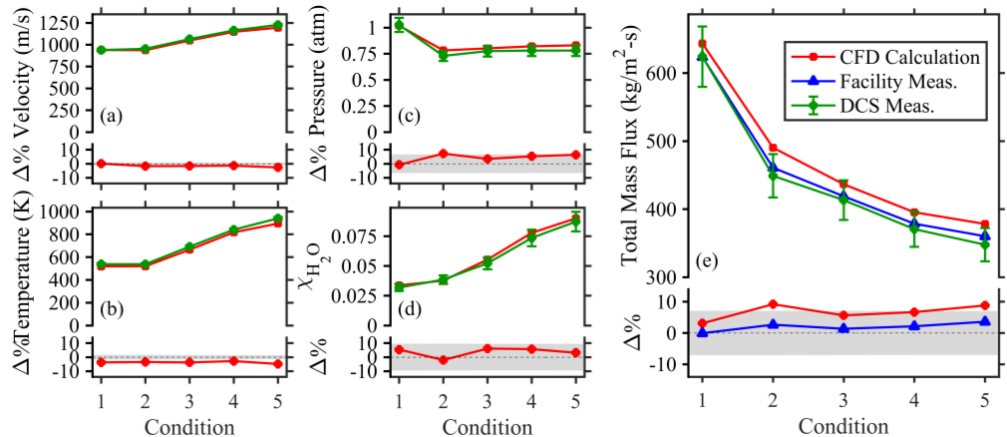

Fig. 4. DCS measurements with uncertainties at the centerline of the isolator for different flow conditions. The DCS fit results (green circles) of velocity (a), pressure (b), temperature (c), $\chi_{H2O}$ (d), and total mass flux (e) are compared to CFD (red squares) with isothermal wall boundary conditions. Each panel has a corresponding residual plot that shows the difference between DCS and CFD values, defined as (CFD-DCS)/DCS, with a shaded gray area denoting the DCS measurement uncertainty. Panel e) also shows a comparison to the facility-derived mass flux values (blue triangles).

*4.2.    Vertical scan*

In the second measurement configuration, we scan the crossed beams vertically from the centerline to within 1 mm of the bottom wall while the isolator flow is maintained at one condition. Figure 5 shows the measurement configuration. The measurements span 7 cm in the axial (streamwise) direction as the beams traverse the isolator at $\theta = 35°, -35°$. The span along the vertical axis is limited to the beam width, which is ~1 mm. Thus the axial resolution is 7 cm and the vertical resolution is 1 mm. Narrower laser beam angles would improve the axial resolution at the expense of velocimetry precision (not accuracy). Figure 5 also shows the CFD-predicted mass flux in the measurement region. One can see that the flow is mostly uniform in the axial direction and that the largest gradient occurs in the vertical direction near the wall. Therefore, we take measurements at 19 mm (centerline) and 9 mm to represent the core flow, and measurements at 3 mm, 2 mm, and 1 mm to profile the boundary layer near the floor (cowl) of the isolator.

Figure 6 shows the DCS vertical profile measurements of the different parameters and the mass flux. The measurements agree with expectations for the various parameters. Velocity is uniform in the core of the flow and decreases at the boundary layer, which is approximately 3-5 mm thick at the bottom of the isolator at this location. Temperature increases near the wall, as expected for supersonic flows where slowing of the gas near the walls increases the static temperature of the gases. Pressure is relatively uniform across the isolator, which we expect for the simple flow conditions used for these experiments (no distortion generator to produce a pronounced oblique shock train in this region of the isolator). Remaining subtle variations in pressure could be due to weak shock disturbances in the flow. Finally, $\chi_{H2O}$ is relatively stable across the isolator, which is expected since water vapor is neither produced nor destroyed in the isolator or near the boundaries.

Figure 6 also incorporates the CFD-derived values of the various parameters. For these experiments, we provide two CFD calculations – one assuming an adiabatic boundary condition at the walls and one assuming an isothermal boundary condition at the walls. The difference between the CFD calculations is primarily visible at the boundary layer. All parameters from both CFD simulations are within 14% of the DCS-measured values. In general, the CFD simulation with adiabatic boundary conditions tends to agree better than the isothermal condition in the boundary layer. Similarly, the adiabatic air mass flux values have better agreement than the isothermal values in the boundary layer. Thus, spatially resolved DCS data can provide an important benchmark and means for tuning CFD in hypersonic flows.

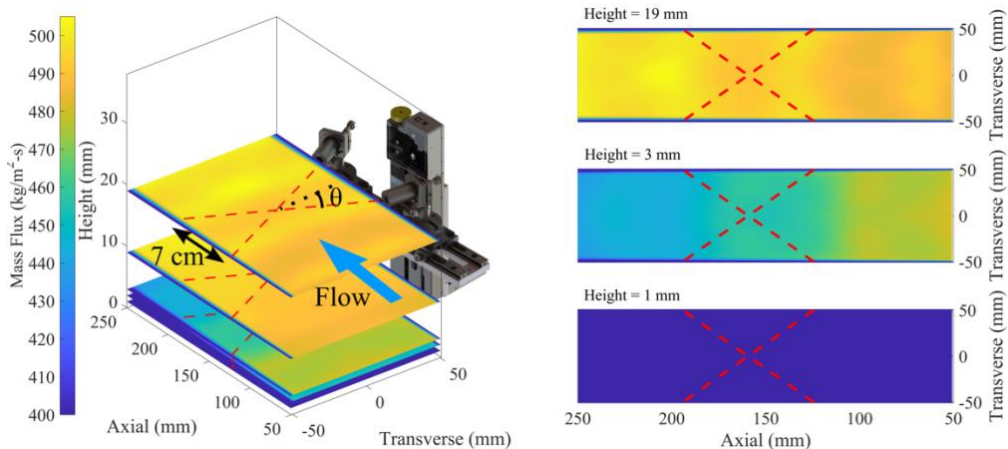

Fig. 5. DCS measurement configuration for the vertical scan experiments. As shown on the left side, measurements are taken at different heights across the isolator at constant flow condition by moving optics with a translation stage. Slices of the CFD-predicted mass flux are included at each height to show the expected gradients in the measurement region. The slices for heights 1 mm, 3 mm, and 9 mm are shown separately on the right for ease of viewing.

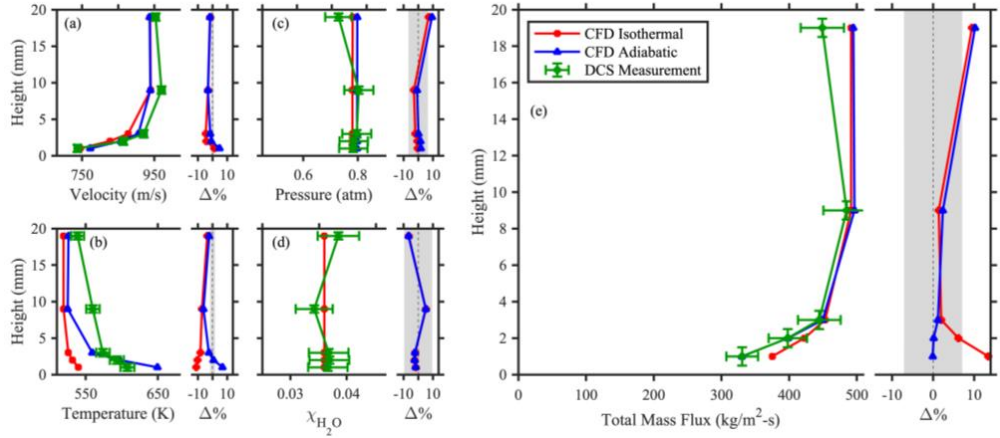

Fig. 6. DCS fit results (green circle) of velocity (a), pressure (b), temperature (c), and $\chi_{H2O}$ (d) and calculated total mass flux (e) compared to CFD for the vertical scan of the lower half of the isolator. Here, both CFD assuming isothermal boundary condition (red circle) is shown alongside CFD assuming adiabatic boundary condition (blue triangle). Differences between DCS and CFD, which is defined as (CFD-DCS)/DCS, is shown to the right of each of the DCS-CFD comparison plots with a grey shaded area indicating the DCS measurement uncertainty. The same color and marker code is used to differentiate which CFD is being compared to. The y-axis uncertainty bars for DCS measurements stem from the width of the laser beam.

## 5. Uncertainty analysis

The DCS instrument is capable of low uncertainty measurements of multiple flow parameters, as described earlier. However, it is important to consider all potential sources of uncertainty to estimate the total uncertainty of the measurements. We identify the sources of uncertainty to be the DCS instrument accuracy, DCS instrument precision, beam angle, background subtraction, and database. The contribution of each source is described in detail in Supplement 1 Sec. 3. Table 1 summarizes the uncertainties stemming from each of these sources for each of the parameters.

The DCS instrument accuracy corresponds to uncertainty stemming from the DCS instrument itself (accuracy and stability of the frequency point spacing and the bandwidth of the system as discussed in Section 2). The DCS instrument precision is the scatter in results due to the noise in the spectrum at 60s of averaging time (absorbance noise $\approx$ 0.003). We find the scatter by calculating the Allan deviation as shown for velocity in Figure 7. The beam angle uncertainty of $\pm 0.25°$ directly affects velocity as it is essential to relating the Doppler shift to velocity. It also introduces a small difference for other parameters due to a slight uncertainty in the laser pathlength through the flow. Uncertainties due to background subtraction arise from differences in background absorption measured before and after the study measurements. The absorption database adds uncertainty due to error in the absorption model parameters (linestrength, pressure broadening and shift) that form the absorption models used to interpret the measured spectra. Finally, air mass flux uncertainty is calculated by combining the uncertainties for the individual parameters according to Eq. 2.

**Table 1. Uncertainty of DCS measurements in run 5**

| Source/Parameter | Velocity | Pressure | Temperature | $\chi_{H2O}$ | Air Mass Flux |
|---|---|---|---|---|---|
| DCS Instrument Accuracy | 0.00015% | 0.012% | 0.015% | 0.0042% | 0.02% |
| DCS Instrument Precision | 0.13% | 0.22% | 0.19% | 0.32% | 0.33% |
| Beam Angle | 0.60% | 0.06% | 0.01% | 0.04% | 0.60% |
| Background Subtraction | 0.52% | 0.27% | 0.66% | 0.38% | 0.88% |
| Database Error | 0.0% | 6.6% | 1.2% | 9.5% | 6.9% |
| Total | 0.8% | 6.6% | 1.4% | 9.5% | 7.0% |

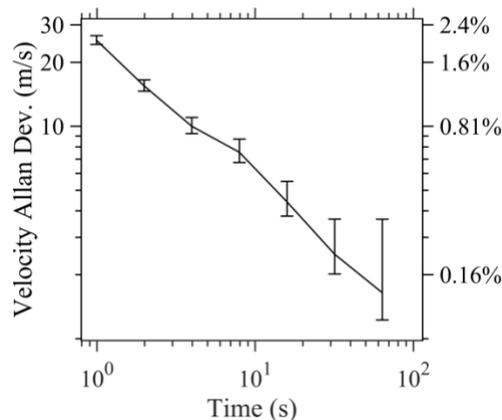

Fig. 7. Allan deviation for velocity from 1s to 60s. Left y-axis tick labels indicate the absolute deviation in m/s and the right y-axis tick labels indicate the relative precision for run 5. Allan deviations for temperature, pressure, and $\chi_{H2O}$ are included in Supplement 1 Sec. 3.2.

Non-DCS sources, especially the absorption database, contribute most of the total uncertainty. We can reasonably expect that uncertainty from non-DCS sources can be improved with better angle determination methods, more vigorous background water vapor purging methods, and database improvement. Thus, overall uncertainty in DCS mass flux measurements can be improved for future measurements.

Uncertainty directly from the DCS instrument (the DCS instrument accuracy and precision) is only ~0.4% at 60s of averaging across all measurements. As the primary uncertainty contribution is from precision, this can be improved with longer averaging times, as the Allan deviation shown in Figure 7 is still declining at 60 seconds. Notably, the velocity precision is <2 m/s at 60 second averaging. Even at a shorter averaging time of 1 second, we achieve a precision of ~2% (25 m/s), indicating the potential for precise, time-resolved measurements.

## 6. Discussion

Despite their importance to many atmospheric and industrial applications, direct measurements of mass flux are extremely challenging in large-scale open environments and environments with extreme flow phenomena because it is difficult to simultaneously and non-intrusively measure all of the components of mass flux (i.e. velocity, temperature, pressure, and mixture composition). Here, we demonstrate non-intrusive, absolute measurements of mass flux in such environments by leveraging the unique properties of stabilized, mode-locked frequency combs. In particular, we demonstrate the measurement technique in a supersonic flow environment where the non-intrusive nature of the sensor survives the high temperatures and flow velocities, and the spatial resolution can help with the presence of variable pressure, shock waves, and boundary layers.

Other laser absorption techniques have been used for a variety of prior measurements in aerospace environments. TDLAS has been used to measure velocity, temperature, or other individual components of mass flux in shock tubes, wind tunnels, and ramjet test models with high resolution, fast acquisition data [49–58]. Lyle *et al.* [61] measured mass flow in a cold, subsonic turbofan aeroengine inlet using $O_2$-absorption TDLAS. Assuming $O_2$ mole fraction is fixed, the density of the air at cold temperatures is determined from absorption feature absorbance area using a room temperature calibration. Chang et al. [47] and Brown et al. [63] made mass flux measurements in supersonic flows. These studies relied on pressure data from wall pressure transducers. Brown et al. [48] also measured the pressure with the laser sensor and found a 32% deviation in mass flux values when incorporating direct pressure measurements instead of the facility-measured pressure value. There are broadband techniques such as super-continuum lasers that allow for broad bandwidth and high acquisition speeds [64]. Studies from Werblinski *et al.* [65,66] demonstrated 10 kHz super-continuum measurements of pressure, temperature and water mole fraction in a rapid compression machine. However, the resolution of the fast supercontinuum measurements (~1 cm$^{-1}$) is broad and not always well characterized compared to the small Doppler shifts (0.01-0.02 cm$^{-1}$) and the absorption feature widths (~0.1 cm$^{-1}$ FWHM) at the conditions in this work. To the best of our knowledge, this is the first study to measure all of the components of mass flux in a supersonic environment with well characterized, low uncertainty.

The ability to localize the pressure, temperature, mole fraction, velocity, and mass flux to the line-of-sight of the laser beam is promising for environments with spatially varying conditions, such as among the boundary layers and pressure shocks that occur in and around hypersonic vehicles. The vertical scan results presented in this paper demonstrate the capability of DCS to measure the spatial variation in the mass flux near the wall of the scramjet isolator. Future measurements could traverse oblique shock trains, incorporating local laser-based pressure measurements as done in this study rather than potentially non-representative facility wall pressure measurements.

The ability to scan a line-of-sight mass flux measurement through spatially varying conditions can be broadened to other open-area measurements of interest. For example, DCS has been demonstrated for long-distance measurements of trace gases [24,26,28]. The technique shown here could be used to measure line-of-sight averaged mass flux conditions downwind of trace gas sources, such as leaking oil and gas production basins, chemical plants, cities, or thawing permafrost. It could also be used with swept beam paths to provide area-averaged mass flux conditions. The large area mass flux approach could have advantages over extrapolating area-representative conditions from point-based flux sampling, such as eddy-covariance [67,2,68] for comparisons where representation error is a concern [69–71]. Additionally, laser array tomographic techniques [58,72] can be employed to potentially provide both spatial resolution and averaging of 2D areas.

Overall, this demonstration of DCS mass flux sensing with low uncertainty across a difficult suite of flow conditions suggest that DCS is an accurate and multi-faceted sensor with applications across many different environments.

**Funding.** Defense Advanced Research Projects Agency (W31P4Q-15-1-0011); Air Force Office of Scientific Research (FA9550-17-1-0224, FA9550-20-1-0328); Air Force Research Laboratories (FA8650-20-2-2418)

**Acknowledgments.** We thank Amanda Makowiecki for advice performing spectroscopic fits using cepstral analysis and Robbie Wright for building the DCS system and for troubleshooting advice during the campaign. We thank Steve Enneking, Andrew Baron, Justin Stewart, and the facility operators who worked hard to ensure a smooth experimental campaign in Research Cell 18. This document has been cleared by the U.S. Air Force under clearance # AFRL-2022-1066.

**Disclosures.** The authors declare no conflicts of interest.

**Data availability.** Data underlying the results presented in this paper are not publicly available at this time as the data contained in the paper has not yet been cleared for public release by the U.S. Air Force.

**Supplemental document.** See Supplement 1 for supporting content.